\documentstyle[12pt]{article}
\advance\textheight by \topskip
\begin{document}
\input epsf

\begin{flushright}
UCLA/96/TEP/8\\
CERN-TH/96-53\\
SU-ITP-96-8\\
hep-th/9602136\\
February 25, 1996\\
\end{flushright}
\vspace{.5cm}

\begin{center}
\baselineskip=16pt

{\Large\bf  SUPERSYMMETRY AND  ATTRACTORS}  \\

\vskip 2cm
 {\bf Sergio Ferrara}\footnote{E-mail:
ferraras@cernvm.cern.ch}\\
 \vskip 0.2cm
Theory Division, CERN, 1211 Geneva 23, Switzerland\\
UCLA Physics Department, Los Angeles CA, USA\\
\vskip .6cm

{\bf Renata Kallosh}\footnote {E-mail:
kallosh@physics.stanford.edu}\\
 \vskip 0.2cm
Physics Department, Stanford University, Stanford   CA 94305-4060, USA\\
\vskip .6cm

\end{center}

\vskip 1 cm
\centerline{\bf ABSTRACT}
\vspace{-0.3cm}
\begin{quote}

We find a general principle which allows one to compute the area of the
horizon of N=2 extremal black holes as an extremum of the central charge.
One considers the  ADM mass equal to the central charge  as a function of electric and magnetic charges and moduli and  extremizes this function in the moduli space (a minimum corresponds to a fixed point of attraction). The extremal value of the  square of the central charge provides the area of the horizon, which depends only on  electric and magnetic charges. 

The doubling of unbroken supersymmetry at the fixed point of attraction for N=2 black holes near the horizon is derived via  conformal flatness of the Bertotti-Robinson-type geometry. These results provide an explicit model independent expression
for the macroscopic Bekenstein-Hawking entropy of N=2 black holes which is manifestly duality invariant. The presence of hypermultiplets in the solution does not affect the area formula. Various examples of the general formula are displayed. We outline  the attractor mechanism in N=4,8 supersymmetries and the relation to the N=2 case. The entropy-area formula in five dimensions, recently discussed in the literature, is also seen to be obtained by extremizing the 5d  central charge.

\end{quote}

\normalsize
\newpage

\normalsize

\section{Introduction}

Supersymmetry seems to be   related to dynamical systems with fixed points describing  the equilibrium and stability\footnote {A point  $ x_{\rm fix}$  where the phase velocity $v(x_{\rm fix})$ is vanishing
is named a {\it fixed point} and represents the system in equilibrium,
$v(x_{\rm fix})=0 $.
The fixed point is said to be an {\it attractor}  of some motion 
$x(t)$ if 
$
\lim_{t  \rightarrow \infty} x(t) = x_{\rm fix}(t).
$}.
The particular property of  the long-term behavior of dynamical flows in dissipative systems is the following: in approaching the attractors  the orbits  lose practically all memory of their initial conditions, even though the dynamics is strictly deterministic. 

The first known to us example of such attractor behavior in the supersymmetric system was discovered in the context of N=2  extremal black holes \cite{FKS,S}. The corresponding motion describes the behavior  of the moduli fields as they approach the core of the black hole. They evolve according to a damped geodesic equation 
(see eq. (20) in \cite{FKS}) until they run into the fixed point near the black hole horizon. The moduli at fixed points were shown to be given  as ratios of charges
in the pure magnetic case \cite{FKS}. Recently Strominger has further shown that
this phenomenon extends to the generic case when both electric and magnetic
charges are present \cite{S}.
The inverse distance to the horizon plays the role of the evolution parameter in the corresponding attractor. By the time moduli  reach the horizon they lose completely the information about the initial conditions, i.e. about  their values far away from the black hole, which correspond to the values of various coupling constants, see Fig.1. 

The main result of this paper is the derivation of the universal property of  the stable fixed point of the supersymmetric attractors: fixed point  
is defined by the new {\it principle of a  minimal central charge}\footnote{We are assuming that the extremum is a minimum, as it can be explicitly verified in some models. However for the time being we cannot exclude situations with different extrema or even where the equation $D_i Z =0$ has no solutions.} and the area of the horizon  is proportional to   the square of the central charge, computed at the point where it is extremized in the moduli space.
In N=2,  d=4 theories, which is the main object of our study in this paper,  the extremization has to be performed in the moduli space of the special geometry and  is illustrated in  Fig. 1. This results in the following formula for the Bekenstein-Hawking entropy $S$, which is proportional to the quarter of the area of the horizon:
\begin{equation}
S={A\over 4}  =  \pi |Z_{\rm fix}|^2  \ , \qquad d=4 \ .
\end{equation}

This result allows generalization for higher dimensions, for example, in five-dimensional space-time one has
\begin{equation}
S={A\over 4}  \sim   |Z_{\rm fix}|^{3/2}    \ , \qquad d=5 \ .
 \end{equation}
  
There exists a  beautiful phenomenon in the black hole physics: according to the no-hair theorem, there is a limited number of parameters\footnote{This number can be quite large, e.g. for N=8 supersymmetry one can have 56 charges and 70 moduli.}  which describe space  and physical fields far away from the black hole. In application to the  recently studied black holes in string theory,   these parameters include the mass, the electric and magnetic charges, and the asymptotic values of the scalar fields. 

 It appears that  for supersymmetric black holes one can prove a new, stronger version of the no-hair theorem:    black holes lose all their scalar hair near the horizon. Black hole solutions near the horizon are characterized only by those discrete parameters which correspond to conserved charges associated with gauge symmetries, but not by the values of the scalar fields at infinity which may change continuously.

A   simple example of this attractor mechanism is given by the dilatonic black holes of the heterotic string theory \cite{G,US}, see Sec. 4 for details. The modulus of the central charge in question  which is equal to the ADM mass  is given by the formula
\begin{equation}
M_{ADM} =|Z| =  {1\over 2}  ({\mbox{e}}^{-\phi_0}  |p| +  {\mbox{e}}^{\phi_0}  |q|) \ .
\label{rec}\end{equation}
In application to this case the general theory, developed in this paper gives the following  recipe to get the area

i) Find the extremum of the  modulus of the central charge as a function of a dilaton ${\mbox{e}}^{2\phi_0}=g^2$ at fixed charges
\begin{equation}
{\partial \over \partial g} |Z| (g, p,q) =   {1\over 2} {\partial \over \partial g}\left( \frac{ 1}{g}  |p| + g  |q| \right) = -\frac{ 1}{g^2}   |p| +   |q| =0 .
\end{equation}

ii) Get the fixed value of the moduli
\begin{equation}
g_{\rm fix}^2=  {\left |p\over q \right |}  \ .
\end{equation}

iii) Insert the fixed value into your  central charge formula (\ref{rec}), get the fixed value of the  central charge:  the square of it is proportional to the area of the horizon 
and defines the Bekenstein-Hawking entropy
\begin{equation}
S = {A\over 4}  =  \pi |Z_{\rm fix}|^2 =  \pi |pq| \ .
\end{equation}
This indeed coincides with the result obtained before by completely different methods \cite{US,Entr}.

In general supersymmetric   N=2 black holes have an ADM  mass $M$ depending on  charges $(p, q)$ as well as on moduli $z$ through the holomorphic symplectic sections  $\left( X^\Lambda (z), F_\Lambda (z)\right)$,
see Appendix.
The moduli present the values of the scalar fields of the theory far away from the black hole. The general formula for the mass of the state with one half of unbroken supersymmetry of  N=2  supergravity interacting with vector multiplets as well as with hypermultiplets is \cite{Cer}-\cite{Ser1}
\begin{equation}\label{F1}
M^2=|Z |^2  \ , 
\end{equation}
where the central charge is \cite{Cer} 
\begin{equation}
Z(z, \bar z, q,p) = e^{K(z, \bar z)\over 2} 
(X^\Lambda(z)  q_\Lambda - F_\Lambda(z) \, p^\Lambda)= (L^\Lambda q_\Lambda - M_\Lambda p^\Lambda) \ , 
\end{equation}
 so that
\begin{equation}\label{F2}
M^2_{ADM} =|Z |^2 = M^2_{ADM} (z,\bar z, p, q) \ .
\end{equation} 
The area, however, is only charge  dependent: 
\begin{equation}\label{F3}
A=A(p, q) \ .
\end{equation} 
This happens since the values of the moduli near the horizon are driven to the fixed point  defined by the ratios of the charges.  This mechanism was explained before in \cite{FKS} and \cite{S} on the basis of the conformal gauge formulation of 
N=2 theory \cite{WLP}.

This attractor mechanism is by no means an exclusive  property of only N=2 theory in four dimensions. Our analysis suggests that  it may be a quite  universal phenomenon in any supersymmetric theory. The main purpose of this paper is   to investigate the attractor mechanism in the symplectic covariant form of N=2 theory, 
to analyse the attractors of N=4 and N=8 theory, and to reinterpret them in terms of N=2 theory. 

In this paper we will use the ``coordinate free" formulation of the special geometry
\cite{CAFP,Cer,Ser1} which will allow us to present a symplectic invariant description of the system.
We will be able to show that the unbroken supersymmetry requires the fixed point of attraction  to be defined by the solution of the duality symmetric equation 
\begin{equation}
D_i Z = (\partial _i + {1\over 2}K_i) Z (z,\bar z, p, q) =0 \ ,
\end{equation}
which implies, see Appendix, 
\begin{equation}
{\partial \over \partial z^i}  |Z|=0
\end{equation}
at 
\begin{equation}
Z=Z_{\rm fix}= Z\left( L^\Lambda(p,q), M_\Lambda(p,q),p,q\right) \ .
\end{equation}

 Equation $\partial _i  |Z|=0$    exhibits the {\it minimal area principle}
in the sense that the area is defined by the extremum of the central charge in the moduli space of the special geometry, see Fig. 2 illustrating this point. 
Upon substitution of this extremal values of the moduli into the square of the central charge we get the Bekenstein-Hawking entropy,
\begin{equation}
S= {A\over 4} =  \pi |Z_{\rm fix} |^2 \ .  
\end{equation}
The area of the black hole horizon  has also an interpretation as the mass of the  Bertotti-Robinson universe \cite{BR}  describing the near horizon geometry.
\begin{equation}
A/4\pi  = M^2_{BR} \ .  
\end{equation}
This mass, as different from the ADM mass, depends only on charges since the moduli near the horizon are in their fixed point equilibrium positions,
\begin{equation}
 M^2_{BR} =|Z_{\rm fix}|^2 = M^2_{BR} (p,q) \ .
\end{equation}
Note that in the Einstein-Maxwell system without scalar fields the ADM mass of the extreme supersymmetric black hole simply coincides with the Bertotti-Robinson one, both being functions of charges:
\begin{equation}
M^2_{ADM} (p,q)
=M^2_{BR} (p,q)\ .  
\end{equation}

  We will describe below a near horizon black holes of N=2 supergravity interacting with vector and hyper multiplets. The basic difference from the pure N=2 supergravity solutions comes from the following: the metric near the horizon is of the Bertotti-Robinson type, as before. However, the requirement of unbroken supersymmetry and duality symmetry forces the moduli to become functions of the ratios of charges, i.e. take the fixed point values. We will describe these configurations, show that they provide the restoration of full unbroken N=2 supersymmetry near the horizon. We will call them N=2 attractors, see Sec. 2.
In Sec. 3 we will analyse some of N=4 and N=8 attractors and provide their interpretation from the point of view of N=2 theory. In Sec. 4  examples of  N=4 and N=8 attractors  will be presented 
using in each case the parameters (``attractor variables") which allow to demonstrate explicitly the dependence of the ADM mass on charges as well as on  moduli and the independence of the area on moduli.
In the last section we make   some remarks on the possible developments of ideas of this paper in the context of looking for the general links between the microscopic and macroscopic physics in supersymmetric theories. Appendix contains a short resume of special geometry.

 \section{N=2 Attractor} 

The special role of the Bertotti-Robinson metric
in the context of the solitons in supergravity was explained 
 by  Gibbons  \cite{Gibb}. He suggested to consider
the Bertotti-Robinson (${\cal BR}$)  metric as an alternative,
maximally supersymmetric, vacuum state. The extreme
Reissner-Nordstr\"om     metric spatially interpolates between this
vacuum and the trivial flat one, as one expects from a soliton.

Near the horizon all  N=2  extremal black holes with one half of unbroken supersymmetry restore the complete  N=2  unbroken supersymmetry. This phenomenon
of the doubling of the supersymmetry  near the horizon was discovered   
in the Einstein-Maxwell system in \cite{Gibb}. 
It was explained in  \cite{K} that the manifestation of this doubling of unbroken supersymmetry is the appearance of a covariantly constant on shell  superfield of N=2 supergravity.
In presence of a dilaton this mechanism was studied in 
\cite{KP}. In the context of exact four-dimensional black holes, string theory and conformal theory on the world-sheet the 
${\cal BR}$ space-time was studied in \cite{LS}.
 In more general setting the idea of vacuum interpolation in supergravity via super p-branes was developed in \cite{GT}.

We will show here using the most general supersymmetric system of N=2 supergravity interacting with vector multiplets and hypermultiplets how this doubling of supersymmetry occurs and what is the role of  attractors in this picture.
The supersymmetry transformation for the gravitino, for the gaugino and for the hyperino  are given in the manifestly symplectic covariant formalism \cite{CAFP,Ser1} \footnote{The notation are given in \cite{Ser1}.} in absence of fermions and in absence of gauging as follows:
\begin{eqnarray}
\delta \psi_{A\mu} &=& {\cal D}_\mu \epsilon_A +\epsilon_{AB} T_{\mu\nu}^- \gamma^\nu \epsilon^B \ , \nonumber\\
\delta \lambda^{iA} &=& i\gamma^\mu \partial_\mu z^i \epsilon^A+
{i\over2} {\cal F}^{i-}_{\mu\nu}\gamma^{\mu\nu}\epsilon_B\epsilon^{AB}\ , \nonumber\\  
\delta \zeta_\alpha &=& i \,{\cal U}^{B \beta} _u \partial_\mu  q^u \gamma^\mu \epsilon ^A \epsilon_{AB}
 C_{\alpha \beta} \ ,
\label{ricca}
\end{eqnarray}
where $\lambda^{iA}$, $\psi_{A\mu}$ are the chiral gaugino and gravitino
fields, $\zeta_\alpha$ is a hyperino, $\epsilon_A\ ,\ \epsilon^A$ are the chiral and antichiral
supersymmetry parameters respectively, $\epsilon^{AB}$ is the $SO(2)$
Ricci tensor. The moduli dependent duality invariant combinations of field strength 
$T_{\mu\nu}^- , {\cal F}^{i-}_{\mu\nu}$ are defined by eqs.  (\ref{tnov}), ${\cal U}^{B \beta} _u $ is the quaternionic vielbein \cite{BW}.

Our goal is to find solutions with unbroken N=2 supersymmetry. The first one is a standard flat vacuum: the metric is flat, there are no vector fields, and  all scalar fields in the vector multiplets as well as in the hypermultiplets take arbitrary constant values: 
\begin{equation}
ds^2 = dx^\mu dx^\nu \eta_{\mu\nu}\ , \qquad T_{\mu\nu}^-={\cal F}^{i-}_{\mu\nu}=0\ ,  \qquad z^i = z^i_0 \ ,  \qquad q^u=q^u_0 \ .
\end{equation}
This solves the Killing conditions $\delta \psi_{A\mu}=\delta \lambda^{iA}=\delta \zeta_\alpha=0$
with constant unconstrained values of the supersymmetry parameter $\epsilon_A$.
The unbroken supersymmetry manifests itself in the fact that each non-vanishing scalar field represents the first component of a covariantly constant N=2
superfield for the vector and/or  hyper multiplet, but the supergravity superfield vanishes.

The second solution with unbroken supersymmetry is much more sophisticated. First, let us solve the equations for the gaugino and hyperino by using only a  part of the previous ansatz: 
\begin{equation}
{\cal F}^{i-}_{\mu\nu}=0\ ,  \qquad \partial _\mu z^i =0\ ,  \qquad \partial _\mu q^u=0
 \ .
\end{equation}
The Killing equation for the gravitino is not gauge invariant. We may therefore consider the variation of the gravitino field strength the way it was done in \cite{K, KP}. Our ansatz for the metric will be to use the geometry with the vanishing scalar curvature and Weyl tensor and covariantly constant graviphoton  field strength $T_{\mu\nu}^-  $ : 
\begin{equation}
R= 0\ , \qquad  C_{\mu \nu \lambda \delta}=0 \ ,  \qquad {\cal D}_\lambda  (T^-_{\mu\nu} )=0  \ .
\end{equation}
It was explained in \cite{K, KP} that such configuration corresponds to a covariantly constant 
superfield of N=2 supergravity $W_{\alpha \beta} (x, \theta)$, whose first component is given by a two-component  graviphoton field strength $T
_{\alpha \beta} $. The doubling of supersymmetries near the horizon happens  by the following reason. The algebraic condition for the choice of broken versus unbroken supersymmetry is given in terms of the combination of the Weyl tensor plus or minus  a covariant derivative of the graviphoton field strength, depending on the sign of the charge. However, near the horizon both the Weyl curvature as well as the vector part vanish. Therefore both supersymmetries are restored and we simply have a covariantly constant superfield $W_{\alpha \beta} (x, \theta)$. The new feature of the generic configurations which include vector and hyper multiplets is that in addition to a covariantly constant superfield of supergravity $W_{\alpha \beta} (x, \theta)$, we have covariantly constant superfields, whose first component is given by the scalars of the corresponding multiplets.  However, now as different from the trivial flat vacuum, which admits any values of the scalars, we have to satisfy the consistency conditions for our solution, which requires that the Ricci tensor is defined by the product of graviphoton field strengths,
\begin{equation}
R_{\alpha \beta \alpha' \beta'}^{BR} = T
_{\alpha \beta}  \bar T_{\alpha' \beta'} \ ,
\label{I}\end{equation}
and that the vector multiplet vector field strength vanishes
\begin{equation}
{\cal F}^{i-}_{\mu\nu}=0 \ .
\label{II}\end{equation}
Before analysing these two consistency conditions in terms of  symplectic structures of the theory, let us describe the black hole metric near the horizon.

The explicit form of the metric is taken as a limit near the horizon $r=|\vec x | \rightarrow0$ of the black hole metric
\begin{equation}
ds^2 = -e^{2U} dt^2 + e^{-2U} d\vec x^2 \ ,
\end{equation}
where 
\begin{equation}
\Delta e^{-U} =0 \ .
\end{equation}
We choose 
\begin{equation}
e^{-2U}= {A\over 4\pi |\vec x |^2}=  {M_{BR}^2\over  r^2} \ , 
\end{equation}
where the Bertotti-Robinson  mass is defined by the black hole area of the horizon
\begin{equation}
M_{BR}^2= {A\over 4\pi} \ . 
\end{equation}

We may show that this metric, which is the Bertotti-Robinson metric  
\begin{equation}
ds^2_{BR} = -  {|\vec x |^2\over M_{BR}^2}
dt^2 +  {M_{BR}^2\over  |\vec x |^2} 
 d\vec x^2 \ ,
\end{equation}
is conformally flat in the properly chosen coordinate system.  In spherically symmetric coordinate system
\begin{equation}
ds^2_{BR} = -  {r^2\over M_{BR}^2}
dt^2 +  {M_{BR}^2\over  r^2}
 (dr^2 + r^2 d\Omega) \ .
\end{equation}
After the change of variables  $r=M_{BR}^2/\rho$
and $|\vec x | = M_{BR}^2/|\vec y |$  
the metric becomes obviously conformally flat
\begin{equation}
ds^2_{BR} = -  {M_{BR}^2\over \rho^2 }
dt^2 +  {M_{BR}^2\over  \rho^2}
 (d\rho^2 + \rho^2 d\Omega)  ={M_{BR}^2\over | {\vec y}| ^2 } ( -dt^2 +  
d \vec y^2) \ ,
\end{equation}
which is in agreement  with the vanishing of the Weyl tensor.

Now we are ready to describe our solution in terms of symplectic structures, as defined in \cite{Ser1}. The symplectic structure of
the equations of motion comes by defining the ${\rm Sp}(2n_V+2)$ symplectic
(antiselfdual) vector field strength
$({\cal F} ^{-\Lambda},{\cal G}^-_{\Lambda} ) $.

Two symplectic invariant combinations
 of the symplectic field strength
vectors are:
\begin{eqnarray}
T^-&=& M_\Lambda {\cal F}^{-\Lambda}- L^{\Lambda } {\cal G}_\Lambda^- \nonumber\\
\nonumber\\
{\cal F} ^{-i}&=&G^{i\bar j} ( D_{\bar j} \bar M_\Lambda {\cal F}^{-\Lambda} -   D_{\bar j}
\bar L^\Lambda {\cal G}_\Lambda^-) \ .
\label{tnov}
\end{eqnarray}
The central charge as well as the covariant derivative of the central charge are defined as
follows: 
\begin{equation} 
 Z = -{1\over 2} \int_{S_2} T^- \ , 
\label{tund}
\end{equation}
and  
\begin{equation} 
Z_i \equiv D_i Z    =  -{1\over 2} \int_{S_2} {\cal F}^{+\bar j} G_{i\bar j}\ . 
\label{tdod}
\end{equation}
The central charge, as well as its derivative,  are functions of moduli and electric and magnetic charges.
The objects defined by eqs. (\ref{tnov}) have the physical
meaning of being the (moduli-dependent) vector combinations which appear in
the gravitino and gaugino supersymmetry transformations respectively.
 In the generic point of the moduli space there are two symplectic    invariants homogeneous of degree 2 in electric and magnetic charges \cite{Ser1}:
\begin{eqnarray}
I_1&=& |Z|^2 + |D_i Z|^2 \ ,\nonumber\\
\nonumber\\
I_2&=& |Z|^2 - |D_i Z|^2 \ .
\end{eqnarray}
Note that 
\begin{eqnarray}
I_1&=& I_1(p,q,z,\bar z)=-{1\over 2} P^t {\cal M}({\cal N}) P\ ,
\nonumber\\
I_2&=&I_2(p,q,z,\bar z)=-{1\over 2} P^t {\cal M}({\cal F}) P \ . 
\label{F}
\end{eqnarray}

Here $P=(p,q)$ and ${\cal M}({\cal N})$ is the real symplectic $2n+2 \times 2n+2$ matrix

\begin{equation}
\pmatrix{
A & B \cr
C & D \cr
}
\end{equation}
where 
\begin{eqnarray}
A&=& {\rm Im} {\cal N} + {\rm Re} {\cal N} {\rm Im} {\cal N}^{-1} {\rm Re} {\cal N} \ , \qquad B =- {\rm Re} {\cal N} \,
{\rm Im} {\cal N}^{-1} \nonumber\\
\nonumber\\
C&=&-{\rm Im} {\cal N}^{-1}  {\rm Re} {\cal N} \ ,  \hskip 2.7 cm  D= {\rm Im} {\cal N}^{-1} \ .
\end{eqnarray}
The vector kinetic matrix  ${\cal N}$ is defined in Appendix. The same type of matrix appears
in (\ref{F}) with ${\cal N} \rightarrow{\cal F} =F_{\Lambda \Sigma}$. 
Both ${\cal N}, {\cal F}$ are K\"ahler invariant functions, which means that they depend only on ratios of sections, i.e. only on $t^\Lambda,  f_\Lambda$, see Appendix. 

The unbroken supersymmetry of the near horizon black hole requires the consistency condition (\ref{II}), which is also a statement about the fixed point for the scalars $z^i(r) $ as functions of the distance from the horizon $r$
\begin{equation}
{\partial \over \partial r} \left (z^i(r) \right ) =0 \qquad  \Longrightarrow  \qquad  D_i Z=0 \ .
\label{fixed}\end{equation}
Thus the fixed point is defined due to supersymmetry by the vanishing of the covariant derivative of the central charge. At this point  the critical values of moduli  become functions of charges, and two symplectic invariants become equal to each other:
\begin{equation}
I_{1 \, \rm fix}= I_{2\,  \rm fix} =( |Z|^2)_{D_i Z=0} \equiv |Z_{\rm fix}|^2 \ .
\label{I=II}\end{equation}
The way to explicitly compute the above is by solving in a gauge-invariant fashion eq. (\ref{fixed}),
\begin{equation}
D_{\bar i}\bar Z = D_{\bar i}\bar L^\Lambda q_\Lambda - {\cal N}_{\Lambda\Sigma}D_{\bar i}\bar L^{\bar i} p^\Lambda = 0 \ .
\end{equation}
By contracting with $D_iL^\Sigma G^{i\bar i}$ and using the property
\begin{equation}\label{u1}
D_iL^\Sigma G^{i\bar i} D_{\bar i}\bar L^\Lambda = - {\textstyle {1\over 2}} {{\rm Im}} {({\cal N}^{-1})}^{\Sigma\Lambda} - \bar L^{\Sigma}L^\Lambda 
\end{equation}
we get 
\begin{equation}\label{u2}
2Z \bar L^\Sigma = i p ^\Sigma - {{\rm Im}} {({\cal N}^{-1})}^{\Sigma\Lambda}\, q_\Lambda + {{\rm Im}}  {({\cal N}^{-1})}^{\Sigma\Gamma}\  {{\rm Re}}   {\cal N}_{\Gamma\Delta}\  p^\Delta \ .
\end{equation}
Here we used the fact that $Z = L^\Lambda q_\Lambda - M_\Lambda p^\Lambda$. This finally gives
\begin{equation}\label{u3}
2i\bar Z  L^\Sigma =   p^{\Sigma} +i  ({{\rm Im}}  {\cal N}^{-1}\  {{\rm Re}}  {\cal N}\, p  +  {\rm Im} {\cal N}^{-1} \, q )^\Sigma \ ,
\end{equation}
and
\begin{equation}\label{u4}
2i\bar Z  M_\Sigma =   q_{\Sigma} +i  ({{\rm Im}}  {\cal N} \, p  + {{\rm Re}}  {\cal N}\    {{\rm Im}} {{\cal N}^{-1} }\  {{\rm Re}}\,  {\cal N} \, p - {{\rm Re}} {\cal N}\  {{\rm Im}}  { {\cal N}^{-1} }\, q )_\Sigma\ ,
\end{equation}
so that

\begin{equation}\label{}
p^\Lambda = i(\bar Z L^\Lambda - Z \bar L^\Lambda)\ , \qquad q_\Lambda = i(\bar Z M_\Lambda - Z \bar M_\Lambda) \ .
\end{equation}

From the above equations it is evident that $(p,q)$  determine the sections up to a
(K\"ahler) gauge transformation (which can be fixed setting $L^0 =e^{K/2}$).
Vice versa the fixed point $t^\Lambda$ can only depend on ratios of charges
since the equations  are homogeneous in p,q.

 The first invariant provides an elegant expression of $|Z_{\rm fix}|^2$ which only involves the charges and the vector kinetic matrix at the fixed point ${\cal N}_{\rm fix} ={\cal N}\left ( t^\Lambda_{\rm fix},  \bar t^\Lambda_{\rm fix} , f_{\Lambda\, {\rm fix}}, \bar f_{\Lambda \,{\rm fix}} \right)$.
\begin{equation}
(I_1)_ {\rm fix} =( |Z|^2 + |D_i Z|^2)_{\rm fix} = -{1\over 2} P^t {\cal M}({\cal N} _{\rm fix}) P =
( |Z_{\rm fix}|^2 ) \ .
\label{N}\end{equation}
Indeed eq. (\ref{N}) can be explicitly verified by using eq. (\ref{old}).
For  magnetic solutions
the   area  formula  was derived in \cite{FKS}. This formula presents the area as the function of the zero component of the magnetic charge and of the K\"ahler potential at the fixed point\footnote{In this paper we have a  normalization of charges which is different from \cite{FKS} due to the use of the conventions of \cite{Ser1} and not \cite{WLP}.}.
\begin{equation}\label{old}
 A = \pi (p^0)^2 e^{-K}  \ .
\end{equation}
In the symplectic invariant  formalism we may check that the area formula (\ref{old})
which is valid for the magnetic solutions (or for generic solutions but in a specific gauge only)
indeed can be brought to the symplectic invariant form \cite{FKS}:
\begin{equation}
A = \pi (p^0)^2 e^{-K} =4 \pi ( |Z|^2 + |D_i Z|^2)_{\rm fix}  =
4 \pi ( |Z_{\rm fix}|^2 )= -2 \pi  p^{\Lambda} {\rm Im}  {\cal F}_{\Lambda \Sigma} p^\Sigma \ .  
\end{equation}

One can also check the first consistency condition of unbroken supersymmetry (\ref{I}), which relates the Ricci tensor to the graviphoton.  Using the definition of the central charge in the fixed point we are lead to the formula for the area of the horizon (which is defined via the mass of the Bertotti-Robinson geometry) in the following form
\begin{equation}\label{new}
M_{BR}^2= {A\over 4\pi} = ( |Z|^2)_{D_i Z=0} \  , \qquad S={A\over 4} = 
 \pi M_{BR}^2 \ .
\end{equation}

The new area formula  (\ref{new}) has various advantages  following from manifest symplectic  symmetry. It also implies the  principle of the  minimal mass of the Bertotti-Robinson  universe, which is given by the extremum in the moduli space of the special geometry.
\begin{equation}\label{uu2}
\partial _i M_{BR} =0 \ .
\end{equation}

\section{N=4,8   $\Longleftrightarrow$ N=2 }
Pure N=4 supergravity consists of  N=2 supergravity and one N=2 vector multiplet. This can be regarded as a $SU(2)\times SU(4)$ invariant truncation of N=8. The N=4 theory exists in two formulations, the $SO(4)$ and $SU(4)$. They are related by duality \cite{Cer}, but for our purpose it is important to observe that the first corresponds to a prepotential $F(X) = -iX^0X^1$, while the second has no prepotential and corresponds to a symplectic change of the basis:
\begin{equation}\label{uu3}
\hat X^0 =X^0\ , \qquad \hat F_0 = F_0\ , \qquad \hat X^1 = - F_1\ , \qquad \hat  F_1 = X^1\  .
\end{equation} 
The charges in these two theories are

1) $SO(4)$
\begin{equation}\label{uu4}
p_0\ , \qquad p_1 = p_0 \ {\rm Re}\ t \ , \qquad q_0 =0\ , \qquad q_1 = p_0 \ {\rm Im } \, t\ .
\end{equation} 
The central charge at the fixed point   is $|Z_{\rm fix}|^2 =  p_0^2 {\rm Re}\,t = p_0 p_1$ and is given by the product of the two magnetic charges.

2)  $SU(4)$ 
\begin{equation}\label{uu5}
p_0\ , \qquad p_1 =0\ , \qquad q_0 =p_0 \ {\rm Im } \, t \ , \qquad q_1 =  p_0\ {\rm Re}\,t    \ .
\end{equation}

In these equations  $t = {X^1\over X^0}$. The central charge  at the fixed point is $|Z_{\rm fix}|^2 =  p_0^2\, {\rm Re}\,t = p_0 q_1$ and is given by the product of electric and  magnetic charge.  This is expected for the dilatonic black hole, see next section. 

In what follows we would like to outline some results concerning the attractive behavior of   N=2 theory and  N=8 theory by taking a consistent N=2 reductions  of N=8. 
In this way one can easily obtain N=2 models with  variety of vector and hyper multiplets. 
The particle decomposition of N=8 to N=2 gives fifteen vector multiplets, $n_v=15$, and ten  hyper multiplets, $n_h=10$. Therefore any model will have those numbers as upper bounds for vector and hyper multiplets. To get a consistent truncation one must choose a subgroup  H of $SU(8)$  such that the two residual supersymmetries are H-singlets. The H invariant states will then give a consistent N=2 theory. 
In particular, the scalar field manifold will be a subspace of ${E_7 \over SU(8)}$ of the form
${\cal S}(n_v) \times {\cal Q} (n_h)$, where ${\cal S}(n_v)\ ,  \,  {\cal Q} (n_h)$ are special and quaternionic manifolds of complex and quaternionic dimensions $n_v, n_h$ respectively. 

A convenient way for obtaining such theories is by considering the untwisted moduli of $ {T_6\over Z_N}$ orbifolds with $H=Z_N \subset SU(3)$.
In this way, by considering type II A, II B theories on such orbifolds, one obtains pairs of models related by $c$-map \cite{N=8 to N=2}
\begin{equation}
(n_v^A, \, n_h^A) \ , \qquad  (n_v^B = n_h^A  -1 , \, n_h^B = n_v^A +1 )
\end{equation}
where $n_v^A = h_{11}^0, \, n_h^A= h_{12}^0+1$. Here   $h_{11}^0,\, h_{12}^0$ are Hodge numbers of the untwisted moduli. This implies that $n_v$ can be at most 9 (because $n_h^{\rm max} =10$). The bound is saturated for the ${T_6\over Z_3}$
orbifold for which
\begin{equation}\label{uu7}
 n_v = 9, n_h= 1\ , ~~ {\rm or } ~~  n_v = 0, n_h= 10 
\end{equation}
in type II A, II B respectively. Since the hypermultiplets do not matter at the level of N=2 theory this appears to be the richest example. In the two cases:

1) ${\cal S}(n_v=9)= \frac{SU(3,3)} {SU(3)\times SU(3) \times U(1)}$\ , ${\cal Q}(n_h=1)= \frac{SU(2,1)} {  SU(2) \times U(1)}$\ ,

2) ${\cal Q}(n_h=10)={E_6\over SU(2) \times SU(6)}$ \ .

The N=8 area formula \cite{KallE(7)}  is a square-root of a quartic invariant constructed out of 56 $Z_{AB}$  central charges, under N=2 reductions $SU(8)\Longrightarrow SU(2) \times SU(6)$ we get 
$$Z_{AB} \Longrightarrow (1,1) + (2,6) + (1,15) \ ,$$
so that the $SU(2)$ invariant part is $(Z, Z_i)$. $Z$ is the N=2 central charge and all $Z_i = D_i Z$  vanish at the fixed point. In this way we necessarily get 
$$A\sim |Z|^2 \ ,$$
 as expected.  Indeed, working out a couple of models and examples, which are consistent truncation of N=8, $SU(8)$ supergravity, we reproduced the result given by the $E_7$ invariant formula \cite{KallE(7)}.  
In the first example we expect to recover the N=8 formula as a function of 10 electric and 10 magnetic charges. We will derive this formula from special geometry in the case where two electric and two magnetic charges exists.  Also we will set to zero the electric and magnetic charges of the other six $U(1)$ gauge fields. This corresponds to a submanifold
 $\frac{SU(1,1)}{U(1)} \times \frac {0(2,2)}{0(2)\times 0(2)} $ in    $\frac{SU(3,3)} {SU(3)\times SU(3) \times U(1)}$.

The appropriate parametrization for a symplectic section in a covariant $O(2,2)$ basis \cite{Cer} is 
$$(X^\Lambda\ , F_\Lambda= S X^\Lambda), \qquad X_\Lambda= \eta_{\Lambda \Sigma} X_\Sigma \ ,$$ 
where $\eta_{\Lambda \Sigma} = (+ +, - -)$ is a Lorentz metric of $O(2,2)$,
$$X^\Lambda X_\Lambda=0\ , \qquad K=-\ln i(S-\bar S)- \ln X^\Lambda \bar X_\Lambda \ .$$
We choose the gauge $X^0=1, X^\Lambda = t^\Lambda $ and $p^\Lambda = p_0 ({\rm Re} X^\Lambda ), q_\Lambda = p_0 ({\rm Re} F_\Lambda)$, moreover we choose $t^1, t^3$ imaginary and $t^2$ real.
The fixed point value of the central charge becomes $|Z_{\rm fix}|^2 = - p_0^2 \, {\rm Im} S (1- ({\rm Re}\,  t^2)^2) $. 
This finally can be reduced to 
\begin{equation}
A\sim \sqrt {|(p_0^2- p_2^2) (q_3^2-q_1^2)|}= \sqrt{|m_0 m_2 e_1 e_3|} \ ,
\end{equation}
where we set 
$m_0= p_0-p_2, m_2= p_2+p_0, e_1= q_3-q_1, e_3 = q_3+ q_1$
This gives the area formula for the solutions found in \cite{cvet,R,KallE(7)} and described in
appropriate (attractor) variables in the next section.

\section{Examples of  N=4,8  attractors}

As already mentioned in the Introduction, we will present here the well known black hole solutions of N=4,8 theories for the convenience of the reader but we will do it  using the adequate variables so that the mass depends on moduli whereas the area obviously does not. To the best of our understanding, this form did not appeared before, neither for N=4 nor for N=8 case. We will call these variables ``attractor variables".

 N=4  dilaton  dyonic black holes \cite{G,US}  near the horizon give an  example of a stable  attractor.
We follow here the description of the black holes near the horizon in \cite{KP}. All notations (up to the $\sqrt 2$ factors) are those of \cite{US}.
The  action we will use is the part of the $SO(4)$ version of the
 N=4,  d = 4 supergravity action without axion,
 \begin{equation}\label{so4action}
 I =\frac{1}{16\pi} \int d^4x\,\sqrt{-g} \Biggl[ -R +
2\, \partial^\mu \phi \cdot \partial_\mu \phi- {1\over 2} \left(
{\mbox{e}}^{-2\phi}F^{\mu\nu}F_{\mu\nu} + {\mbox{e}}^ {2\phi}\tilde
G^{\mu\nu}\tilde G_{\mu\nu}\right) \Biggr]\ ,
\end{equation}
where the $SO(4)$ field $\tilde G_{\mu\nu}$ is related to the
$SU(4)$  field $G_{\mu\nu}$ as follows
 \begin{equation}\label{dual}
\tilde G^{\mu\nu} = {\frac{i}{2}}
\frac {1}{\sqrt{-g}}\, e^{-2\phi}\,
\epsilon^{\mu\nu\lambda\delta}\,
G_{\lambda\delta}\ . \end{equation}
This means that each time we have an electric $SO(4)$ field, it correspond to the magnetic $SU(4)$ one and vise versa.
For extreme  supersymmetric dilatonic black holes, the fields are
built out of
two functions $H_1$ and $H_2$ \cite{US} :
 \begin{eqnarray}
ds^{2} &=&  e^{2U}dt^{2}-e^{-2U}d\vec{x}^{2} \ ,\nonumber\\
A = \psi dt  &, &\qquad \tilde B = \chi dt  \ , \nonumber\\
F =  d \psi \wedge dt &, &\qquad  \tilde G = d \chi \wedge dt \ ,
\nonumber\\
e^{-2U} = H_1 H_2  &, &\qquad
e^{2\phi} = H_2/ H_1  \ ,\nonumber\\
\psi = \pm H_1^{-1} &,&\qquad  \chi
=\pm H_2^{-1} \ ,
\label{anz}\end{eqnarray}
where the condition on the functions $H_1,H_2$ is that they be
harmonic,
\begin{equation} \label{Hcond}
\partial_i\partial_i H_1 =0\ , \qquad \partial_i\partial_i H_2=0 \,
{}. \end{equation}
We use isotropic coordinates $r^2 = \vec x^2 $ and we define, as different from \cite{KP} and \cite{US} 
\begin{equation}
H_1 = {\mbox{e}}^{-\phi_0} +\frac{ |q|}{r} \ , \qquad
H_2 = {\mbox{e}}^{+\phi_0}+\frac{ |p|}{r} \ .
\label{bh}\end{equation}
The metric becomes
\begin{equation}
g_{tt}^{-1} = g_{ii}  = e^{-2U} = \left ({\mbox{e}}^{-\phi_0} + \frac{ |q|}{r}\right) \, \left ({\mbox{e}}^{+\phi_0}+\frac{ |p|}{r}\right )= 1 +   \frac{ {\mbox{e}}^{-\phi_0}  |p| + {\mbox{e}}^{\phi_0}   |q|}{r} + { |pq| \over r^2}\ .
\end{equation}
The dilaton is 
\begin{equation}
 e^{-2\phi} ={  {\mbox{e}}^{-\phi_0} + \frac{|q|}{r} \over   {\mbox{e}}^{+\phi_0}+\frac{|p|}{r}} \ .
\end{equation}
This explains everything: the mass defined by the $1/r$ term in this expression when $r\rightarrow \infty$ depends on charges and moduli, whereas the area, defined by the $1/r^2$ term when $r\rightarrow 0$, depends only on the charges $p$ and $q$.
The mass $M$ and the dilaton charge $\Sigma$ are related to the $U(1)$
electric $q$
and magnetic $p$ charges as
\begin{equation}
M =  {\textstyle{1\over 2}}({\mbox{e}}^{-\phi_0}  |p| +  {\mbox{e}}^{\phi_0}  |q|) \ , \qquad  \Sigma = {\textstyle{1\over 2}}({\mbox{e}}^{-\phi_0}  |p| -  {\mbox{e}}^{\phi_0}  |q|) \, \, .
\end{equation}
Thus the black hole solution is characterized by three independent parameters: two charges $p,q$ and the value of the dilaton at infinity ${\mbox{e}}^{-\phi_0}$. In particular, the mass of the black hole depends on all three parameters.
We will now find that the black hole solution near the horizon is described completely
by the two charges: the value of the dilaton at infinity becomes irrelevant. No matter what was the value of the dilaton ${\mbox{e}}^{-\phi_0}$ at infinity, near the horizon it 
is driven to the fixed point given by 
$$({\mbox{e}}^{-2 \phi}) _{\rm fix} =    \frac{|q|}{|p|} \ .
$$
Consider  the extreme $pq \not= 0$ dilatonic black holes near the
horizon,
 in the limit $r \rightarrow 0$,  i.e. in the limit $1/r \equiv \rho \rightarrow \infty$.  The metric  in (\ref{anz})
becomes
\begin{equation} ds^2 =
\frac{{r}^2}{|pq|} \,  dt^2 -\frac{|pq|}{r^2}
 \, d r^2 - |pq| \, d\Omega^2  \ . \label{RBtype}
\end{equation}
This metric is precisely the ${\cal BR}$ metric.
 The dilaton for these solutions behaves as
\begin{equation} \label{dilrbl}
{\mbox{e}}^{-2 \phi} =   \frac{|q|}{|p|} \,
 \biggl( 1 +   \frac {{\mbox{e}}^{-\phi_0}  |p| - {\mbox{e}}^{\phi_0} 
|q|}  {  |pq| \rho }  \,  
+ O(1/\rho^2) \biggr) \, ,
\end{equation}
so we see that the term linear in $1/\rho$ is proportional to the
dilaton charge
$\Sigma$.  The electric and magnetic fields are given by
\begin{equation}\label{FGrbl} F =   \frac{1}{
q}  \,
dr
\wedge dt \ ,  \qquad \tilde{G} =   \frac{1}{
p} \,
dr \wedge dt
\,  , \end{equation}
or equivalently, in terms of dual fields
\begin{equation}\label{FGrbl2} \tilde F =  
q  \sin \theta  \,
d\theta  \wedge d\phi 
\ ,  \qquad {G} =   
p\,  \sin \theta  \,
d\theta  \wedge d\phi 
\   . \end{equation}
The dilaton has vanishing derivative at $\rho \rightarrow \infty$, which is a fixed point. The value of the dilaton given in eq. (\ref{dilrbl}) shows that close to the fixed point the dilaton has a positive derivative or a negative derivative depending on the sign of the dilaton charge
$\Sigma$.
An example of a  basin of attraction for the dilaton is given in Fig. 1. Independently of initial conditions for the dilaton at infinity all trajectories are attracted to a fixed point  $({\rm e}^{-2\phi})_{\rm fix}=4$  near $r=0$.

The example of the N=8 attractor is given using the truncated action of N=8 supergravity. The form of this solution is a slight modification of the one obtained in \cite{cvet,CT,Tsey, R,KallE(7)}. The modification makes the area independence of moduli manifest.
\begin{eqnarray}
S&=& {1\over 16\pi G} \int d^4x \sqrt {-g}\, \Bigl( R-{\textstyle{1\over 2}}
\left [(\partial \eta)^2 +
(\partial \sigma)^2 +(\partial \rho)^2
\right]  \nonumber\\
\nonumber\\
&-& {e^{\eta}\over 4}  \left [ e^{\sigma + \rho}  (F_1)^2 +
e^{\sigma - \rho}  (F_2)^2 + e^{-\sigma - \rho}  (F_3)^2 +e^{-\sigma + \rho}
(F_4)^2  \right] \Bigr) \ .
\end{eqnarray}
\begin{eqnarray}
ds^2 & = & -e^{2U} dt^2 +
e^{-2U} dx^2,    \hskip 3 cm e^{4U}=  \psi_1 \psi_3 \chi_2\chi_4  \ ,
\nonumber\\
 \nonumber\\
e^{-2\eta}&=&\frac{\psi_1\psi_3}{\chi_2\chi_4}\ , \, \qquad  \qquad
e^{-2\sigma }=\frac{\psi_1\chi_4}{\chi_2\psi_3}\ , \, \qquad \qquad
e^{-2\rho}=\frac{\psi_1\chi_2}{\psi_3\chi_4}\ , \nonumber\\
 \nonumber\\
F_1&=& \pm d\psi_1 \wedge dt \ ,  \quad
 \tilde F_2= \pm d\chi _1 \wedge dt\ , \quad
F_3= \pm d\psi_3 \wedge dt \ ,  \quad  \tilde F_4 = \pm d\chi _4 \wedge dt \ ,
\end{eqnarray}
where
\begin{eqnarray}
\psi_{1} &=& \left( e^{\eta_0 +\sigma_0 + \rho_0 \over 2} +{|q|_{1} \over r}\right)^{-1}  , \quad \chi_{2} =
\left( e^{-\eta_0 -\sigma_0 + \rho_0 \over 2}+{|p|_{2} \over r_{2}}\right)^{-1}  , \\  \psi_{3} &=& \left(
e^{\eta_0 -\sigma_0 - \rho_0 \over 2}+{|q|_{3} \over r_{3}}\right)^{-1}   , \quad \chi_{4} = \left( e^{-\eta_0 +\sigma_0 - \rho_0 \over 2}+{|p|_{4}
\over r_{4}}\right)^{-1}  ,
\end{eqnarray}
and magnetic potentials correspond to $\tilde{F}_{2/4}=e^{\eta\pm
(\sigma-\rho)}F^*_{2/4}$\, . Here $^*$
denotes the Hodge dual. We may keep in mind the standard definition of the moduli in terms of the constant values of 
$S,T,U$ fields at infinity: 
\begin{equation}
e^{-\eta_0} = {\rm Im} S\ , \qquad e^{-\sigma _0} = {\rm Im} T\ ,  \qquad e^{-\rho_0} = {\rm Im} U \ .
\end{equation}
The metric becomes 
\begin{equation}
g_{tt}^{-1} = g_{ii} = e^{-2U}=  \left (\psi_1 \psi_3 \chi_2\chi_4 \right)^{-1/2}
\end{equation}
At infinity $r\rightarrow \infty$ it is 
\begin{equation}
g_{tt}^{-1} = g_{ii} \rightarrow 1+ {1\over 2r}\Bigl(e^{-\eta_0 -\sigma_0 - \rho_0 \over 2}|q_1| +
e^{\eta_0 +\sigma_0 - \rho_0 \over 2 }|p_2| +e^{-\eta_0 +\sigma_0 + \rho_0 \over 2}|q_3| +e^{\eta_0 -\sigma_0 + \rho_0 \over 2}|p_4|\Bigr) 
+ \dots   \end{equation}
This shows that the mass depends heavily on the values of moduli, in addition to dependence on charges. However, near the horizon $r\rightarrow 0$ we get a nice and simple dependence only on charges:
\begin{equation}
g_{tt}^{-1} = g_{ii} \rightarrow { |q_1\, p_2\,q_3\, p_4|^{1/2} 
\over r ^2 }+ \dots \ ,
  \end{equation}
which defines the properties of the area formula.
The fixed point values of moduli at the attractor $r\rightarrow 0$ are
\begin{equation}
(e^{-2 \eta})_{\rm fix} =\left | \frac{p_2 \, p_4}{q_1 q_3}\right | \ ,
\qquad  (e^{-2\sigma})_{\rm fix} =\left | \frac{p_2 \, q_3}{q_1 p_4}\right| \ , \qquad  (e^{-2\rho})_{\rm fix} =\left | \frac{p_4 \, q_3}{q_1 p_2}\right| \ .
\end{equation} 
The previous  N=4  case  is the special case of this solution  with trivial $T, U$:
\begin{equation}
|q_1|= |q_3|  \ ,  \qquad |p_2|= |p_4| \ .
\end{equation}

\section{Discussion}

In this paper we have found a complete description of N=2\, d=4 attractors which serve to define the entropy-area-central charge formula of the most general extremal black holes in N=2 supergravity interacting with the arbitrary number of  vector and hyper multiplets.

To support our point of view that the extremization of the central charge in the moduli space is the generic phenomenon of any supersymmetric theory describing  non-rotating black holes with the non-vanishing area of the horizon we discuss the extension of the above analysis, in five dimensions, when 5d black holes are considered. Details will be given elsewhere.

In N=1 case (which reduces to N=2 of d=4) the underlying geometry of vector multiplets is real \cite{GTS} (called ``very special geometry" in ref. \cite{WP}). It is defined in terms of symmetric constants $d_{ABC}$ which multiply the geometrical term
\begin{equation}
\omega^{ABC} =  \int A^A \wedge F^B \wedge F^C    \ , \qquad A,B,C = 0, \dots , n_v \ ,
\end{equation}
to build the $d_{ABC}\, \omega^{ABC} $  term in the effective action.
The central charge is \cite{Anton}
\begin{equation}
Z(z, q)= t^A (z) q_A \ , 
\end{equation}
where $t^A$ is subject to the constraint 
\begin{equation}
d_{ABC}  t^A(z)  t^B(z)  t^C(z)  =1 \ .
\end{equation}
Here  $z^i$ denote real coordinate of the $n_v$ dimensional manifold with the metric
\begin{equation}
G_{ij} = -3 \partial_i t^A\, d_{AB} (t)\, \partial_{j} t^B \ , 
\end{equation}
where $d_{AB}(t(z)) \equiv  d_{ABC} t^C(z)$, $(d_A \equiv d_{AB} t^B,\  d_A \partial _i t^A =0$).
Unbroken supersymmetry for the ${\cal BR}$ metric requires, as in d=4,
\begin{equation}
\partial_i Z \left( t(z), q  \right) =0 \ .
\end{equation}
The ${\cal BR}$ mass is then $Z(q)= Z \left(t(z), q \right)|_{\partial_i Z =0}$.
The area is proportional to $Z^{3/2}(q)$ and therefore it is possible to give a general expression at d=5 for N=1 extremal black hole entropy:
\begin{equation}
S={A\over 4} \sim \left (Z(q)\right) ^{3/2} \ , 
\end{equation}
where $Z(q)$ is given below.
This formula  for particular choices of $d_{ABC}$ can be also applied to the  N=2 (N=4 of  d=4) d=5 black holes of type II strings, compactified on $K_3\times S_1$, recently discussed in the literature \cite{ascv,Tsey}. 
The point is that there is a sector in common with the heterotic string compactified on $K_3\times S_1$ \cite{Cadavid} which has N=1 supersymmetry at d=5.
The sector contains three vectors, the dual of $B_{\mu\nu},  B_{\mu 6}, g_{\mu6}$ and in the heterotic case gives two matter vectors and the graviphoton.
By denoting by $e_s,e_1,e_2$ their charges and using the vector parametrization of $Z$ as  in  
ref. \cite{Anton} it is straightforward to show that\footnote{We conjecture in analogy with ref. \cite{KallE(7)} that the cubic expression in (\ref{cubic}) is related to the cubic 
E(6) invariant of the d=5, N=4 theory (N=8 of d=4).}
\begin{equation}
Z|_{\partial_i Z =0} \sim (e_s e_1 e_2 )^{1/3} \ ,
\label{cubic}\end{equation}
and therefore 
\begin{equation}
A\sim Z^{3/2}
|_{\partial_i Z =0} \sim \sqrt {Q_H Q_F^2} \ , 
\end{equation}
where $Q_H= e_s , Q_R\pm Q_L = e_1, e_2, Q_F^2\equiv Q_R^2 - Q_L^2=e_1 e_2 $ in notation of ref. \cite{ascv}.

The formula above
 is a particular case of a general formula for $Z$ valid for any N=1, d=5 theory  which we report here
\begin{equation}
Z_{\rm fix} =\sqrt {  \left (d^{AB}(q)\right)^{-1} q_A q_B} \  ,  \qquad A \sim \left [ \left (d^{AB}(q)\right)^{-1} q_A q_B \right ]^{3/4}  \ ,
\label{5darea}\end{equation}
 where $\left (d^{AB}(q)\right)^{-1}= \left (d^{AB} (t(z))|_{\partial_i Z  =0}\right)^{-1} $. Equation  (\ref{5darea}) applies in particular to eleven dimensional supergravity compactified on Calabi-Yau threefold.

It would be interesting to find the general class of 5d black holes with 1/2 of unbroken N=1, d=5 (N=2 of  d=4) supersymmetry with the area of the horizon realizing the formula (\ref{5darea}).
\vskip 0.7 cm

It was emphasized over the years by Susskind \cite{Suss} that the  Bekenstein-Hawking  entropy of a ground state of a system is a logarithm of a number of microstates of string theory. Therefore it cannot vary continuously  and should depend on charges since the charges are discrete and not continuous parameters. This idea become particularly appealing  from the time that the entropy of $U(1)^2$ dilatonic black holes was shown   to be proportional to the product of charges $PQ$ in $U(1)^2$ theory \cite{US}  and 
to $\sqrt{P_1Q_2 P_3Q_4}$ in $U(1)^4$ case \cite{cvet}. 
This idea was studied and  further developed  in   \cite{lawi, CT}. 

The important property of the entropy of supersymmetric black holes was proved in \cite{US}: 
according to supersymmetric non-renormalization theorem 
the entropy does not change when quantum corrections are taken into account  (in theories where there are no  supersymmetry anomalies).  The basic reason for   the supersymmetric non-renormalization theorem comes from the fact that the unbroken supersymmetry of the bosonic configuration is associated with the fermionic isometries in the superspace. Using Berezin's integration rules over anticommuting variables one can show the absence
of quantum corrections to the effective euclidean on shell action related to  the entropy.  

Quite recently  a  dramatic progress was achieved in the understanding the microstates of the string theory, which has allowed to  compare the macroscopic and the microscopic calculation of the entropy \cite{ascv} -\cite{spin}. This again at the moment goes from one striking example to another. The most recent review of the known dyonic extremal black holes with the non-vanishing area can be found in \cite{Tsey}.

We believe that the general property of  extremization of the central charge in the moduli space which was found in this paper in the context of the four-dimensional N=2 supergravity and static extremal black holes 
may be generalized for higher supersymmetries, higher dimensions (like we have shown it in d=5 case) and rotating stationary  black holes. It may become a  universal principle, which will control the value of the area of the horizon and the Bekenstein-Hawking entropy of the extreme black holes and other extreme objects with the non-vanishing area of the horizon.  This principle  may finally make it possible to not merely  accumulate various examples of amazing things  happening with supersymmetric solitons but serve as a  link between macroscopic  and microscopic  systems including black holes, strings, p-branes and d-branes.

\section*{Acknowledgements}
We are grateful to  Andrew Strominger for having shown us his paper \cite{S}  prior to publication. We appreciate stimulating conversations  with F. Larsen,  A. Linde, L. Susskind and E. Witten. We would like also  to express our gratitude to Arvind Rajaraman who  asked us if it is possible to  explain the Strominger-Vafa  5d area formula and its apparent asymmetry in charges from the point of view of  central charge extremization. 

S.F. was  supported in part
by DOE under grant DE-FGO3-91ER40662, Task C and by EEC Science program 
SC1*ct92-0789 and INFN.
 R.K. was  supported
by  NSF grant PHY-9219345.
\newpage

\section*{Appendix: Resume on special geometry}
Symplectic sections are defined as 
\begin{equation}
(L^\Lambda, M_\Lambda) ,\qquad \Lambda = 0,1,...n,
\end{equation}
where $(L,M)$ obey the symplectic constraint
\begin{equation}
i(\bar L^\Lambda M_\Lambda - L^\Lambda \bar M_\Lambda)=1 \ .
\end{equation}
 $L^\Lambda(z, \bar z) $   and $M_\Lambda(z, \bar z)$ depend on $z,\bar z$, which are the coordinates of the ``moduli space".  Special geometry relations are 
\begin{eqnarray}
M_\Lambda &=& {\cal N}_{ \Lambda \Sigma} L^\Sigma \ , \nonumber\\
D_{\bar i} \bar  M_\Lambda &=& {\cal N} _{\Lambda \Sigma} D_{\bar i}  \bar L^\Sigma \ .
\end{eqnarray}
$ L^\Lambda$ and $M_\Lambda$ are covariantly 
holomorphic (with respect to K\"ahler connection), e.g. 
\begin{equation}
D_{\bar k} L^\lambda = (\partial_{\bar k} - {1\over 2} K_{\bar k})L^\Lambda =0 \ .
\end{equation}
This equation can be solved by setting 
\begin{equation}
L^\Lambda = e^{K/2} X^\Lambda \ , M_\Lambda = e^{K/2} F_\Lambda\ , \qquad (\partial_{\bar k}
X^\Lambda  = \partial_{\bar k}
F_\Lambda=0) \ .
\end{equation}
The K\"ahler potential is 
\begin{equation}
K = -\ln i( \bar X^\Lambda F_\Lambda -   X^ \Lambda  \bar F_\Lambda) \ ,
\end{equation}
and the K\"ahler  metric $G_{i\bar i} = \partial_i \partial_ {\bar i} K$ with the inverse metric $G_{i\bar i}^{-1} =G^{i\bar i}$.

It is obvious that the ratios
\begin{equation}
t^\Lambda = {L^\Lambda\over L^0} = {X^\Lambda\over X^0} 
\end{equation}
are holomorphic in the coordinates and gauge invariant
\begin{equation}
\partial_{\bar k} t^\Lambda (z, \bar z) =0, \qquad   t^\Lambda= t^\Lambda (z) \ .
\end{equation}
Consider now the quadratic matrix 
\begin{equation}
e^a_i(z) = \partial_i t^a (z) \ , \qquad a=1,\dots n, \qquad t^0 =1 \ .
\end{equation}
If  $e^a_i$ is invertible we can choose a  frame where 
\begin{equation}
e^a_i(z) = \delta^a_i \ , ~~~  t^a = z^i \delta_i^a \ ,
\end{equation}
i.e. the sections $t^\Lambda$ can be identified with the moduli coordinates ($t^\Lambda$ are called special coordinates). In this frame one can further show that $F_\Lambda$ is integrable, i.e.  $F_\Lambda= \partial_\Lambda F$,
and that $F(X)=( X^0)^2 f(t)$ and ${1\over X^0} \partial_\Lambda F =\left  ( { \partial \over \partial t_a} f(t) , f_0 (t) = 2f(t) - t^a \partial_a f(t) \right )$. Since $|L_0|= e^{K/2} |X^0|$, by a K\"ahler transformation  $X^\Lambda \rightarrow X^\Lambda e^{-f(t)}$  we can set $X^0=1$ and get $|L_0|= e^{K/2}$ as in the conformal gauge of \cite{WLP}.
If  $e^a_i$ is not invertible no prepotential exists in the chosen symplectic basis. This is what happens in some examples of Sec. 3.

Note that $X^\Lambda(z)$ are subject to holomorphic redefinitions (sections of a holomorphic line bundle):
\begin{equation}\label{xx1}
X^\Lambda(z) \to X^\Lambda(z)~e^{-f(z)} \ ,
\end{equation}
so that \begin{equation}\label{xx2}
L^\Lambda(z) \to  L^\Lambda(z)~e^{\bar f(z) -f(z)\over 2}\  .
\end{equation}
This occurs because $L^\Lambda = e^{K/2} X^\Lambda$ and $K\rightarrow K+f+\bar f$
under  K\"ahler transformations, so that 
 \begin{equation}\label{xx3}
Z(q,p,z) \to Z(q,p,z)~e^{\bar f(z) -f(z)\over 2} \ .
\end{equation}
We will show  in what follows that $D_iZ = 0$ implies $\partial_i|Z| = 0$. 

Generically $\bar Z$ is covariantly holomorphic:
 $D_i\bar Z  =  (\partial_i - {\textstyle{1\over 2}} K_i) \bar Z =0$,  which leads to $\partial_i \bar Z = {\textstyle{1\over 2}} K_i \bar Z$, however 
$D_i  Z  =  (\partial_i + {\textstyle{1\over 2}} K_i)  Z \neq 0$. Only at the fixed point we have to satisfy the constraint 
 $D_iZ = 0$,  that  implies $D_iZ \bar Z= 0$, $D_i(Z\bar Z) = 0$.
It follows that
$D_iZ \bar Z + ZD_i  \bar Z = (\partial_i + {\textstyle{1\over 2}} K_i)  Z \bar Z + Z (\partial_i - {\textstyle{1\over 2}} K_i) \bar  Z = \partial_iZ\bar Z + Z\partial_i \bar Z= \partial_i|Z|^2 =2 |Z|\partial_i|Z|$=0. 

 $|Z|$ is both symplectic and K\"ahler gauge invariant, this is why the connection drops  and $D_i Z = 0$ ($D_{\bar i}Z = 0$) entails $\partial_i |Z| = 0$.

\vfill
\newpage

\begin{figure}
\centerline{ \epsfxsize 4in\epsfbox{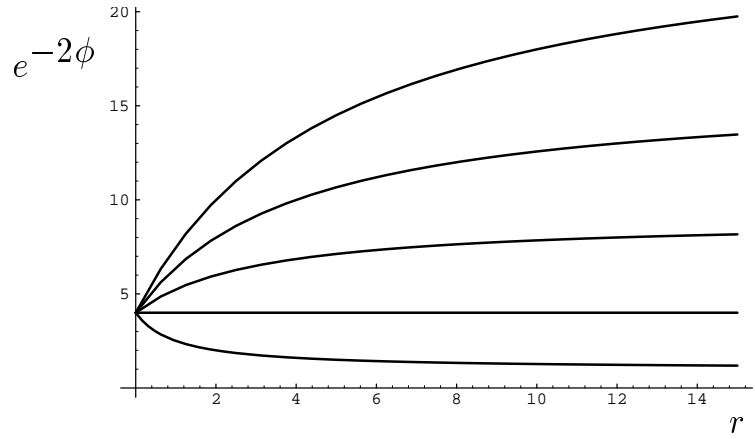}}
 \vskip 1.5cm
\caption{Evolution of the dilaton from various initial conditions at infinity to a common fixed point
at $r=0$.}

\label{F22}

\end{figure}

\begin{figure}

\centerline{\hskip 1 cm\epsfxsize 3.5in \epsfbox{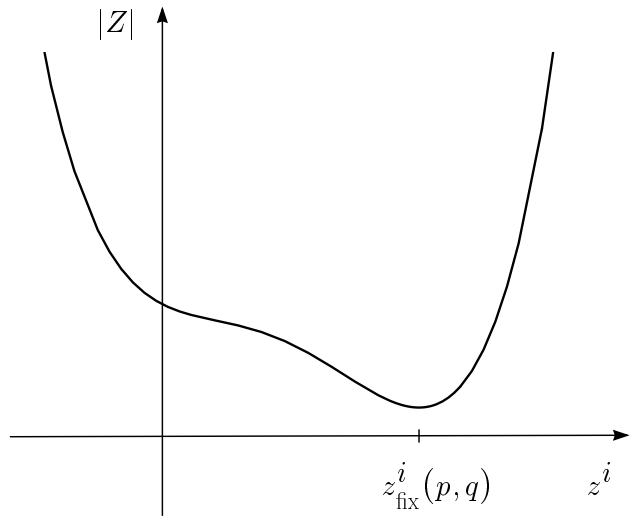}}
 \vskip 2 cm
\caption{Extremum of the central charge in the moduli space.}

\label{F11}

\end{figure}

\end{document}